\documentclass[12pt]{article}
\usepackage{color,amsmath,amsfonts,amssymb,a4,epsfig}

\newcommand{\R}{{\mathbb{R}}}
\newcommand{\C}{{\mathbb{C}}}

\newcommand{\E}{{\mathbb{E}}}

\def\pa{\partial}
\def\ra{\rightarrow}

\def\gd{\delta}
\def\ge{\varepsilon}

\def\gg{\gamma}

\def\gl{\lambda}
\def\go{\omega}

\def\gs{\sigma}

\def\OPD{${\rm \Psi}$DO}

\newtheorem{defi}{Definition}[section]

\newtheorem{lemm}{Lemma}[section]

\newtheorem{coro}{Corollary}[section]
\newtheorem{theo}{Theorem}[section]
\newtheorem{exem}{Example}[section]

\begin{document}

\title{A Semi-classical calculus of  correlations\footnote{to appear
in the thematic issue 
``Imaging and Monitoring with Seismic Noise''
  of  the series  ``Comptes Rendus G\'eosciences'',
from  the  Acad\'emie des  sciences} }

\author{Yves Colin de Verdi\`ere \footnote{Institut Fourier,
 Unit{\'e} mixte
 de recherche CNRS-UJF 5582,
 BP 74, 38402-Saint Martin d'H\`eres Cedex (France);
\textcolor{blue}{\tt http://www-fourier.ujf-grenoble.fr/$\sim $ycolver/}}}

\maketitle
\begin{abstract}
{\bf Fran\c{c}ais.}

 La m\'ethode d'imagerie passive en sismologie 
 a \'et\'e d\'evelopp\'ee  r\'ecemment en vue
d'imager la cro\^ute terrestre \`a  partir d'enregistrements du bruit
sismique.
 Elle repose sur le calcul des fonctions   de
 corr\'elation de  ce bruit. Nous donnons dans cet article des 
formules explicites pour cette corr\'elation dans le r\'egime
``semi-classique''.
Pour cela, nous d\'efinissons le spectre de puissance d'un champ
al\'eatoire comme l'esp\'erance de sa mesure de Wigner, ce qui permet
d'utiliser un calcul dans l'espace des phases : le calcul
pseudo-diff\'erentiel
et la th\'eorie des ``rays''.
Nous obtenons ainsi une formule pour la corr\'elation du bruit sismique
dans le r\'egime ``semi-classique'' avec une source de bruit qui
peut \^etre localis\'ee et non homog\`ene.

Nous montrons ensuite comment l'utilisation des ondes guid\'ees de
surface
permet d'imager la cro\^ute terrestre.

{\it Mots cl\'es} : Imagerie passive ; semi-classique ;
ondes de surfaces. 

{\bf  English.}

The method of passive imaging in seismology has been developped 
recently
in order to image the earth crust from recordings of the seismic
noise.
This method is founded on the computation of correlations of the
seismic noise.
In this paper, we give an explicit formula for this correlation in the
``semi-classical'' regime.
In order to do that, we define the power spectrum of a random field
as the ensemble average of its Wigner measure, this allows phase-space
computations: the pseudo-differential calculus and the ray theory.
This way, we get a formula for the correlation of the seismic noise 
in the semi-classcial regime with a source noise which can be localized
and non homogeneous.
After that,  we show how the use of surface guided waves allows
to image the earth crust. 

{\it Keywords} : passive imaging, semi-classics,
surface waves. 
\end{abstract}

\section*{Introduction}

Correlations of the noisy wave fields is used 
as a new tool in seismic imaging and monitoring,
 starting from the pioneering work
of  Campillo  and Paul \cite{C-P} (similar tools have been used 
in helio-seismology \cite{DJ})  and followed by many works
\cite{DL,DLC,RSK,SRTDHK,SS,SC,SCSR,WL3,We}.
See also the review paper \cite{GS}. It has also been used in the monitoring
of the deformations of volcan{\oe}s \cite{Br}.
Because it is a very powerful method and, hopefully,  in order to make it more 
efficient, it is  quite 
challenging to give mathematical supports to this method, now called 
``passive imaging''.
This has been done in a rather great generality 
 in \cite{YCdV1,YCdV2} using semi-classical analysis
(see also \cite{LW,RSK,BGP,GS}).

Exact formulas for the correlations of the fields are known if
the source noise is homogeneous (a white noise).
This assumption is not satisfied in applications. It is therefore
desirable to get formulate valid for more general source noises, in
particular
if the source noise is localized in some part of the domain.
This turns out to be possible in the so-called semi-classical regime
where the wave-lengths are negligible with respect to
the size of the propagation domain. 
The field correlation admits a general expression in terms
of the Green's function and the source correlation (Equation
(\ref{equ:general})).
The idea is to find the asymptotics of this expression
in the semi-classical regime.

I will present in this paper approximate formulas which
are valid in the  range of high frequency wave propagation
and for which the source noise is  localized in
some part of the domain of propagation.
The correlation is explicitly given in term of the decomposition
of the Green's function as a sum over rays and the (phase-space)
power
spectrum of the source noise.
I can   use ray theory  if I assume that the 
source noise has a short correlation distance of the same order
of magnitude than the wavelengths.
 This calculus  can be presented   in a very
geometric way using rays propagation  as well
as a re-interpretation of the source correlation in terms   
of the phase space power spectra.  
I  use  the calculus of pseudo-differential
operators in a very essential way.
I will not reproduce  the  mathematical arguments which
are presented in my paper \cite{YCdV2},   but I will try
not only to give  explicit  formulas, but also  to present 
the main ideas and tools.

Here is a more precise description of the content:
the goal is to get the formula given in Theorem \ref{theo:main}
which gives the modification of the correlation of the seismic
noise induced by the non-homogeneity of the source
noise. The modification is given in terms of the power spectrum
of the source noise, the attenuation and the ray dynamics associated
to the deterministic wave equation.

I first give a review of the pseudo-differential calculus
(section \ref{sec:opd}): this allows to put the basic 
terminology of rays dynamics and 
to define power spectra of arbitrary random fields (section \ref{sec:random}).

I then introduce the simplest mathematical model where the source
noise is simply the right-handside of the wave equation (section
 \ref{sec:math-model}) and I present our main formula in section 
\ref{sec:main-formula}.
The interest  of the result depends of the relation between 2 time
scales discussed in section \ref{sec:time}:
the Ehrenfest  time given in terms of the Lyapounov exponent and the
attenuation time.

How to use all of this in imaging problems?  I do that
(section \ref{sec:surface}) in the case
of seismology using the effective wave equation for the  guided surface
waves. The  final problem turns out to be an inverse spectral
problem whose mathematical solution is known.

Finally, I discuss in section \ref{sec:scatt} a
 related issue, namely  the calculus of the
correlations of plane waves scattered by an obstacle
or an inhomogeneity  
viewed as random waves: the direction of the waves is supposed
to be random and uniform. This way, I show that the
  result of   \cite{SS} is completely
general.

\section{A short review of the pseudo-differential calculus
and Wigner measures} \label{sec:opd}

For the mathematics of  pseudo-differential
operators, see  \cite{Di-Sj,Ev-Zw,Fo,Tr}.

  The {\it pseudo-differential operators}
(${\rm  \Psi}$DO's)
 were introduced in the sixties by Calderon, Zygmund, Nirenberg,
 H\"ormander and others 
 as a tool in the study of linear partial differential equations
with non constant coefficients.
They provide also  the geometrical extension of Hamiltonian formalism
of {\it classical mechanics} to {\it wave mechanics} (see \cite{Du}).
 In applications to physics,
it is often called the {\it ray} theory  (see \cite{Po}).
 The same tools apply to the 
study of the semi-classical limit of quantum mechanics and to the high 
frequency  limit of wave equations (acoustic, electromagnetic or
seismic waves).

 There is a {\bf small parameter} $\ge >0$  in the theory
which is the Planck ``constant''  $\hbar  $
in quantum mechanics and 
the wave length or more precisely the dimensionless ratio
between the wave length and the size of the propagation domain
for wave equations. 
Most results are only valid in the limit $\ge \ra 0$, but,  
for simplicity, the reader  can think of $\ge$ as a fixed, small enough,
number.

\subsection{${\rm \Psi}$DO's  }

A  pseudo-differential operator (${\rm \Psi}$DO) ~ on $\R^d$
is a linear operator on functions $f:\R^d \ra \C$,
$A_\ge:={\rm Op}_\ge (a) $, 
 defined using a suitable function defined
on the phase space,  $a(x,\xi):\R^d_x \oplus \R^d_\xi  \ra \C $,
 called the {\it symbol} of $A_\ge$, 
by the formula (Weyl quantization)
\[ A_\ge (f)(x)=\frac{1}{(2\pi)^d}\int_{\R^d \times \R^d} e^{i
  (x-y).\xi
}
a\left(\frac{x+y}{2},\ge \xi \right) f(y) dy d\xi ~.\]
The function $a$   is assumed to be
 smooth
and  homogeneous near  infinity in $\xi$.
The Schwartz kernel\footnote{The ``Schwartz kernel'' of a linear operator
$A$ is the ``continuous matrix'' of $A$, we will denote it by
$[A](x,y)$ and it is characterized
by
$Af(x)=\int_X [A](x,y) f(y) dy $.}
 $[A_\ge ](x,y)$  of $A_\ge$ is located 
near the diagonal $x=y$ and is of the form
\[ [A_\ge ](x,y)\cong _{\ge \ra 0} k(x,(x-y)/\ge ) \]
where $k(x,z) $ is 
 a smooth function outside $z=0$  going to $0$ as $z\ra \infty$.

Simple examples are
\begin{itemize}
\item $ {\rm Op}_\ge (1)={\rm Id} $ by the Fourier inversion formula
\item $ {\rm Op}_\ge (\xi _j)= \frac{\ge }{i}\frac{\pa  }{\pa x_j } $
\item$  {\rm Op}_\ge (x_j)$ is the multiplication by $x_j$
\item If $\chi $ is a positive function with bounded
support, the operator
$ {\rm Op}_\ge (\chi (\xi)) $ is a frequency filter
\item  $ {\rm Op}_\hbar ( |\xi|  ^2 +V(x))=
-\hbar^2 \Delta +V(x) $: the  Schr\"odinger operator
\item  $ {\rm Op}_\ge ( n(x) |\xi|  ^2)=-\ge^2 {\rm div} \left(
n(x) ~{\rm grad} \right)$: the acoustic wave operator. 
\end{itemize}

The main properties are the following ones which hold as 
$\ge \ra 0$:
\begin{itemize}
\item Composition:
\[ {\rm Op}_\ge (a)\circ {\rm Op}_\ge (b)\approx 
{\rm Op}_\ge (ab) \]
\item Brackets:
\[  [{\rm Op}_\ge (a), {\rm Op}_\ge (b)]\approx 
\frac{\ge}{i}  {\rm Op}_\ge \{ a, b \} \]
where 
\[ \{ a, b \}= \sum _{j=1}^d \left(
 \frac{\pa a }{\pa \xi_j}\frac{\pa b }{\pa x_j}
-\frac{\pa a }{\pa x_j}\frac{\pa b }{\pa \xi_j} \right) \]
is the Poisson bracket.
This last property is very important because it relates the algebra
of ${\rm \Psi}$DO's to the geometry of the phase space given
by the Poisson bracket. 
\end{itemize}

\subsection{Wigner functions}

Wigner functions define the localization of energy in the phase space
$\R^d_x \times \R^d_\xi $
for a wave function $u=u(x)$.
They involve the  scale $\ge $.
The Wigner function $W_u^\ge (x,\xi)$ of $u$  is the function on the
phase space defined 
by the identities 
\[\forall a\in C_0^\infty \left(\R^{2d}\right),~
 \int_{\R^{2d}} a(x,\xi) W_u^\ge (x,\xi) dx d\xi =
 \langle  {\rm Op}_\ge (a) u|u \rangle~, \]
where $\langle u|v \rangle =\int u(x)\bar{v}(x) dx$,  
or
\[ W_u^\ge(x,\xi)=  \frac{1}{(2\pi  )^d }\int_{\R^d} e^{-iv.\xi }u\left(
  x+\frac{\ge v}{2}
\right)
\bar{u}\left(x-\frac{\ge v}{2}\right) dv  ~.\]

I have
\[ \int_{\R^d} W_u^\ge(x,\xi) d\xi = |u(x)|^2 ,~ 
\int_{\R^d} W_u^\ge(x,\xi) dx = |{\cal F}_\ge u(\xi)|^2  ~,\]
where ${\cal F}_\ge u(\xi)= $
is the $\ge-$Fourier transform of $u$ given by
\[ {\cal F}_\ge u(\xi)=\frac{1}{(2\pi \ge)^{d/2}}\int e^{-i x.\xi 
  /\ge }u(x) dx ~.\] 
This means that the marginals of the Wigner measure
$W_u^\ge (x,\xi)dxd\xi $ are $ |u(x)|^2dx$
  and $ |{\cal F}_\ge u(\xi)|^2d\xi $.

\subsection{Hamiltonian dynamics and ray method}

Let us consider the wave equation $u_{tt} -{\cal L}u=0$
where ${\cal L}$ is an elliptic \OPD ~
like the acoustic operator ${\cal L}={\rm div}(n~{\rm grad})$.
The symbol, usually called the dispersion relation,  of this equation is
$\go^2 -n(x)\|\xi\|^2 =0$.
To this relation
is associated  a dynamics called the ray dynamics given by the
Hamilton
equations: 
\begin{equation}\label{equ:ham}
 \frac{dx_j}{dt}=\frac{\pa H}{\pa \xi_j}, ~
\frac{d\xi_j}{dt}=-\frac{\pa H}{\pa x_j} ~\end{equation}
with $H= \sqrt{n}\| \xi \|$.
The main result (Theorem \ref{theo:main} below)
 uses the ``Hamiltonian flow'' $\Phi_t $:
 $\Phi_t (x,\xi)$ is the
value at time $t$ of the previous differential system (\ref{equ:ham})
with data $(x,\xi)$ at the time $t=0$.
In the case of an homogeneous medium, $n=n_0=$constant,
I have
\[ \Phi_t (x,\xi)=(x+t \sqrt{n_0}\xi/\| \xi \|, \xi )~.\]
In the ray theory, this correspond to the group velocity of waves $\sqrt{n_0}$.

The mathematical theory of rays is called the theory of Fourier
Integral
Operators and has been developed in the seventies by
H\"ormander and Duistermaat (see \cite{Du})
 following some pioneering work of Lax and
Maslov. A presentation more adapted to physicists is given
in \cite{Po}.
 Unfortunately, the geometric background is rather sophisticated
and cannot be presented in a few pages. However, explicit formulas in
terms of oscillatory functions and oscillatory integrals are
available.

In what follows, I will use the fact  that the Green's
function $G(t, x,y)$ of  wave equation  
admits, in the semi-classical regime  (short wave-length),
 a decomposition  as a sum of contributions of rays $\gg$ 
going from $y$ to $x$ in time $t$:
$G=\sum _\gg G_\gg $.

\section{Random fields: power spectra and correlations }
\label{sec:random}

Let $f=f(x),~x\in\R^d,  $ be a random complex-valued  field 
with  zero mean value. 
 Let us denote by $ \E $ the
expectation or ensemble average.

\begin{defi}
 The {\bf correlation} of the random field $f$ is the 2-points
function 
given by  
\[ C(x,y):= \E (f(x)\bar{f}(y) ) \]

 The {\bf power spectrum}  of the random field $f$
is the  function on the phase space given by  the expectation  
of the Wigner functions
\[ p_\ge  :=\E ( W_f^\ge  ) ~.\]
\end{defi}
The power spectrum and the correlation contain the same information:
\begin{itemize}
\item The correlation 
$  C(x,y)$  is  ($(2\pi \ge )^d $ times)  the operator kernel of
$ { Op}_\ge (p)$ or
\[ C(x,y)=\int e^{i\langle x-y |\xi \rangle /\ge }p_\ge \left(
\frac{x+y}{2},\xi \right) d\xi ~.\]
\item 
$p_\ge $  is  ($(2\pi \ge )^{-d} $ times) the symbol of the operator
whose integral kernel is  $C$.

\end{itemize}

\begin{exem}:
 the {\bf  white noise}
\end{exem}
 $C=\gd (x-y)$, $p_\ge=1 /(2\pi \ge )^d $.
\begin{exem}:
 a {\bf   stationary noise} on $\R$ with $\ge=1$,
$C(s,t)=F(s-t)$ and
$p_1(s,\omega)$ is the Fourier transform $ {\cal F}(F)(\go) $. 
\end{exem}

\section{A mathematical model} \label{sec:math-model}

I will now discuss the basic mathematical model: it consists of 2 
parts :
\begin{itemize}
\item A {\it deterministic wave equation} which could be the elastic wave
equation or more simply here the acoustic wave equation.
Because the source of noise will be permanent, some attenuation
in the equation is needed.
\item A {\it source noise} assumed to be stationary and ergodic in time.
In seismology, this source is usually  created by the interaction
of the fluids surrounding the earth crust (atmosphere or ocean)
with the crust itself.
This source is modeled by a random field which  I put on 
the right-handside of the equation. 
\end{itemize}

For simplicity, I will discuss only  the case of a scalar acoustic
wave
equation on some domain in $\R^d_x$
 with a random source field $f=f(x,t)$ ($t$ is the time):

\begin{equation}\label{equ:wave}  u_{tt} + a(x) u_t - {\cal L} u =f
 \end{equation}
where 
\begin{itemize}
\item The field $u=u(x,t)$ is scalar
\item $a $, the attenuation,  is a smooth  $>0 $ function.
I  will assume  for simplicity  that $a$ is time independent, but it 
is not really necessary 
\item ${\cal L} $ is a self-adjoint
pseudo-differential operator of symbol $-\ge ^{-2}l_0^2 (x,\xi)$.
Usually, $l_0$ is homogeneous of degree $1$ which makes 
 ${\cal L} $ independent of $\ge$. This will not be the case
for dispersive
waves like surface waves.  
Typical
examples are 
 the  Laplace-Beltrami operator of a Riemannian metric on
  $X$ with $l_0(x,\xi)=\sqrt{ g^{ij}(x)\xi_i \xi_j }$
and    the acoustic wave operator
${\rm div}\left(n(x)~{\rm grad}\right)$)
 with 
 $l_0 (x,\xi)=
\sqrt{ n(x)}\|\xi \|$.
I introduce $L_0:= {\rm Op}_\ge (l_0)=\ge \sqrt{-{\cal L}}$ 
\item $f=f(x,t)$ is a  stationary  and  ergodic (in time)
random field with correlation
$\E (f(s,x)f(s',y))= \gd (s-s') \Gamma (x,y)$
 and  power spectrum $p(x,\xi)  $; I assume that $p(x,\xi)$
 has bounded support and  that $f$
is real valued and hence that $p(x,\xi)$
 is even w.r. to $\xi$.
I assume that $p$ is independent of $\ge$, this implies that 
the correlation is $\ge-$ dependent: in particular,
$ \Gamma (x,y)<< |x-y|/\ge $. The source noise  decorrelates rapidly
as $|x-y| >> \ge$. 
\end{itemize}

The Green's function is the integral kernel $G$
giving the causal solution of 
Equation (\ref{equ:wave}) in terms of $f$:
\[ u(x,t)=\int _0^\infty ds \int _X  G(s,x,y)f(t-s,y)dy ~.\]

Our goal is to compute the correlation
\[ C_{A,B}(\tau )=
\lim _{T\ra + \infty }
\frac{1}{T} \int_0^T  u(A,t)u(B,t-\tau ) dt ~.\]
\begin{lemm}
The following relation holds:
$C_{A,B}(-\tau)=C_{B,A}( \tau )$.
\end{lemm}
Hence I can (and will!)  restrict ourselves  to $ \tau >0$.

Using the fact that the source noise is ergodic and
stationary, I get the following result
\begin{theo} The field correlation is given by Equation
  (\ref{equ:general}) only 
in terms of the Green's function $G$ and the correlation
$\Gamma $  of the source
noise
\begin{equation} \label{equ:general}
 C_{A,B}(\tau )= ~\int_0^\infty ds \int _{X\times X}
dxdy G(s+\tau, A,x) G(s,B,y)\Gamma (x,y) .\end{equation}
\end{theo}
All the work is now concentrated to get a more explicit  and more 
geometric expression: this will be done using an expression of the
Green's
function as a sum over rays going from $B$ to $A$ in time $\tau $
and using the power spectrum $p$ of $f$ which is a semi-classical expression
of the correlation of the source noise.

\section{The main formula} \label{sec:main-formula}


Let us denote by $\Omega _\pm (t)$ the ``one-parameter groups''
of linear operators 
generated by $\pm i L_0 -\ge a/2 $:
$\Omega _+ (t) u_0 $ is the solution 
of the differential equation
$\dot{u}= (\frac{i}{\ge} L_0 -a/2 )u$
with $u(0)=u_0$, and similarly for $\Omega _-(t)$.
The use of  $\Omega _\pm (t)$ is a way to split the Green function of the
wave equation usually given by some ``sinus'' function into 2 
exponentials: this way, I reduce the wave equation from an equation
with of second order in time to a diagonal system of first order
in time.  
\[ \Omega _\pm (t)= e^{t\left( \pm \frac{i}{\ge} L_0 -a/2 \right) }=
e^{t\left( \pm {i}{\sqrt{-{\cal L}}} -a/2 \right) } ~.\]
I will express the result in terms of operators instead of expressing
them in terms of their kernels (matrices). This gives a much more
compact
expression! 
The symbol $\circ$ means the composition of operators
while $\hat{C}(\tau )$ is the operator
whose integral kernel (matrix) is $C_{A,B}(\tau )$:
\[ (\hat{C}(\tau) u )(A)=\int _X C_{A,B}(\tau )u(B) dB ~.\]

The main result is
\begin{theo} \label{theo:main}
The correlation is given, for $\tau >0$, as $\ge $ goes to $0$,   by
\begin{equation} \label{equ:corr-wave}
  \hat{C}(\tau )\cong\left[ \Omega _+ (\tau )
+ \Omega _- (\tau )\right] \circ \Pi~,\end{equation}
with $\Pi = {\rm Op}_\ge (\overline{\pi} ) $
and
\begin{equation} \label{equ:pi} \overline{\pi }(x,\xi) =
\frac{\ge^2}{4 l_0^2}\int_{-\infty}^0 
e^{-\int_t^0 a( \Phi _s (x,\xi) ) ds}p ( \Phi_t (x,\xi) ) dt ~,\end{equation}
and if $a= a_0 $ is constant
\[ \overline{\pi }(x,\xi) =\frac{\ge^2}{4 l_0^2}\int_{-\infty}^0 
e^{-a_0 |t|}p ( \Phi_t (x,\xi) ) dt ~.\]
\end{theo}

I will compare our result (Equations (\ref{equ:corr-wave}) and
(\ref{equ:pi}))  to the Green's function.

In the semi-classical regime, i.e. as $\ge \ra 0$, I have 
\[ G(t,A,B)\cong \frac{\ge }{2i}   \left[(\Omega _+( t)-\Omega _-(t))
  \circ L_0^{-1} \right] (A,B)~,\]

Let us now compute the $\tau-$derivative of $C_{A,B}(\tau)$:
\[ \frac{d}{d\tau } C(\tau)\cong  -
\frac{\ge }{i}\left( \Omega _+ (\tau) -\Omega _- (\tau)
\right)\circ L_0  \circ \Pi  ~.\]
In the case of white noise and constant attenuation $a_0$,
 I know (see for example \cite{YCdV2} Section 5.1  for a derivation)) that 
\begin{equation} \label{equ:white-noise}
  \frac{d}{d\tau } C(\tau)= -\frac{1}{2a_0}G(\tau ) ~\end{equation}
which is consistent 
with the previous semi-classical formula.

I can now give a more concrete formula:
\begin{coro}
Writing $G(\tau, A,B)$ as a sum $\sum_\gg  G_\gg $ of contributions of
rays
$\gg (s)$  with $\gg (0)=(B,\xi_B) $ and
$\gg (\tau )=\Phi_\tau (B,\xi_B)=(A,\xi_A)$,
I get
\[  \frac{d}{d\tau } C(\tau,A,B)\cong
\sum_\gg  M_\gg G_\gg ~,\]
with
\[ M_\gg =  -\frac{1}{2} \int _{-\infty}^0 
e^{-\int_t^0 a(\gg (s))  ds}p (\gg (t)  ) dt ~.\]
\end{coro} 
In the case of the white noise $p=1$ and $a=a_0$, I recover
the  formula
\begin{equation} \label{equ:white-noise2}
 M_\gg= -1/2a_0 ~. \end{equation} 

Let us also remark that, if there is an unique trajectory
from
$B$ to $A$ in time $\tau $, the prefactor $M_\gg$ applies to
the Green's function itself. It is the case, if I work
with wave equations with constant coefficients in $\R^n$.

The previous formula is consistent with the observations of the
paper \cite{SCS}: the correlation $C_{A,B}(\tau) $
is not always an even function of $\tau $ as it is if the source is 
a white noise. The evenness is valid only up  to scaling of
$C_{A,B}(\tau) $:
\[ C_{B,A}(\tau)=C_{A,B}(-\tau)\sim k C_{A,B}(\tau )~.\]
The factor $k$ is the ratio of the integrals giving
$M_\gg$ for the ray $\gg(t)$ going from $B$ to $A$ and
$\gg(-t)$ going from $A$ to $B$.

\section{Time scales} \label{sec:time}

As I see from the general expression of the correlation
given in Equation (\ref{equ:general}), the proof of the 
main theorem \ref{theo:main} involves the 
knowledge of the Green's function at large times.
This is a well known difficulty and the semi-classical
expansions of the Green's functions are valid up to to
the so-called {\it Lyapounov  time} which involves the Lyapounov 
exponent measuring the rate of  instability of the
ray dynamics.
Roughly speaking, the Lyapounov exponent is the smallest number
$\lambda $ so that the distance between any  to rays
$\gamma _1(t)$ and $\gamma _2 (t) $ satisfies the estimates 
\[
d(\gamma _1(t),\gamma _2(t))\leq 
Ce^{\gl t }d(\gamma _1(0),\gamma _2(0)~\]
with $C$ independent of $\gamma _1(0)$ and $\gamma _2(0)$.
There is an associated time scale $T_{\rm Lyap} =1/\lambda $.
On the other hand there is an attenuation time scale  for
the wave dynamics expressed in terms of the decay of 
the Green's function 
\[ |G(t,x,y)| \leq C e^{-T/T_{\rm att}}~.\]
$T_{\rm att} $ satisfies the estimate
$T_{\rm att} \geq 2/ \inf a $.
The approximation given in Theorem \ref{theo:main}
is better when $T_{\rm att} >> T_{\rm Lyap} $.
In particular, this condition is necessary in order to get
point-wise convergence (i.e. convergence for $A$ and $B$
fixed).

\section{The use of surface waves for passive imaging}
\label{sec:surface}

A remarkable  application of the previous tool is  to the imaging
of the earth crust \cite{LW,WL1,WL2,C-P,SC,SCSR}.
 This is done using the part of the Green's
function associated to the surface waves: the earth crust acts as a
wave guide on elastic waves and these waves follow an effective wave
equation. The effective Hamiltonian is described now:
let us start with the acoustic wave equation
$ u_{tt} - {\rm div}( n ~{\rm grad} u) =0 $
with the function $n$ coming from a stratified medium
$n=n({\bf x}, z )$ (here $z=0$ is the surface)
 where $n$ is weakly dependent of ${\bf x}$ 
(this can be formalized as $n({\bf x},z)= N(\ge {\bf x},z)$
with $N$ smooth and $\ge $ small). 
Using the  adiabatic separation of variables
$u\sim U (\ge x,z)e^{i\langle {\bf x} |\xi \rangle } $
with $U$ weakly dependent of ${\bf x}$, I can operate
as if $n$ was  independent of ${\bf x}$ 
 and I get
the reduced equation
\[ U_{tt} + {\rm Op}_1(\gl ({\bf x},\xi)) U =0 \]
where $\gl ({\bf x},\xi)$ is an eigenvalue of the  Sturm-Liouville
operator
\[ L_{{\bf x},\xi}=-\frac{d}{dz}n({\bf x},z) \frac{d}{dz}+n({\bf
  x},z)\| \xi \|^2 \]
with appropriate boundary conditions at $z=0$.

From the correlation, I get the ray dynamics of the surface
waves and hence the effective
Hamiltonians $\gl ({\bf x},\xi)$.
 The inverse problem to be solved is 
the following inverse spectral problem:
from  the fundamental mode (or any other available mode) of $ L_{{\bf x}_0,\xi}$ 
in some range of wave numbers $|\xi|$, recover $n( {\bf x}_0,z)$. 
This is the kind of well posed inverse problem for which
analytical/numerical
method can be used (see \cite{YCdV5}).

\section{A formula for the scattering of 
random plane waves} \label{sec:scatt}

I have seen an exact formula for the correlation
of the wave field when the attenuation $a$ is constant and 
the source noise is a white noise. 
I will see another exact formula in the context of wave scattering
by a perturbation sitting in a bounded domain of $\R^d$
(see \cite{YCdV1}).
This formula is very general and applies in all situations of
wave scattering (scalar or elastic waves),
 i.e. for any medium  which is homogeneous near infinity:
non-homogeneity's lies  at finite distances or there
is a   scattering by a bounded obstacle. 
This calculus was motivated by the result of \cite{SS}, showing
that this result   is completely general.

Let us consider for example  an acoustic wave equation (\ref{equ:wave})
with $n=n_0$ outside a bounded set of $\R^d$.
I will consider scattering solutions of the stationary
wave equation
\begin{equation} \label{equ:helm} {\rm div }(n ~{\rm grad} u) -\go^2 u=0
\end{equation}
which are of the following form:
let us define, for ${\bf k} \in \R^d$,  the plane wave
\[ e_0 (x,{\bf k})=e^{i{\bf k}.x }~.\]
I am looking for solutions  
\[ e (x,{\bf k})= e_0 (x,{\bf k})+ e^s(x,{\bf k})\]
of  
 equation (\ref{equ:helm})
 in $\R^d $, with $n_0k^2=\go^2$(\footnote{As often, I denote
$k:=|{\bf k}|$ and $\hat{{\bf k}}:={{\bf k}}/{k}$}), 
 where $e^s$, the scattered wave, satisfies the so-called
Sommerfeld radiation condition:
\[  e^s(x,{\bf k})=\frac{e^{ik|x|}}{|x|^{(d-1)/2}}\left( e^{\infty}
\left(\frac{x}{|x|},{\bf k }\right) + O\left(\frac{1}{|x|}\right) \right),
~x\ra \infty ~.\]
The complex function $e^{\infty}
(\hat{x},{\bf k })$ is usually called the 
{\it scattering amplitude} and is a signature of the inhomogeneities.
The functions  $e (x,{\bf k})$ are deformed exponentials and allow
to write an explicit spectral decomposition of our wave operator,
which is a ``deformation'' of the Fourier transform.

Let us look at $e(x,{\bf k})$ as a random wave with
$k=\go/\sqrt{n_0}$ fixed.
The point-point correlation of such a random wave
 $C_\go ^{\rm scatt}(x,y)$
is given by:
\[  C_\go ^{\rm scatt}(x,y)=\int _{k\sqrt{n_0}=\go }
 e(x,{\bf k})
\overline{e(y,{\bf k})} d\gs (\hat{k}). \]
 
It is proved in \cite{YCdV1}, section 8,  that 
\[  C_\go ^{\rm scatt}(x,y)=-\frac{2^{d+1}  \pi ^{d-1}
 n_0^{d/2} }{ \go ^{d-2} }
\Im (G(\omega  +i0,x,y))~,\]
where $G(\omega,  x,y)$ is the stationary Green's function, i.e.
the Schwartz kernel of $\left(\omega ^2 +{\rm div}(n~{\rm grad })\right)^{-1}$.

\section{Conclusions} \label{sec:concl}

I hope to have convinced the reader,
even if he is not very much involved in mathematics,
 that it is possible to derive
rather explicit asymptotic formulas for the correlation 
$C_{A,B}(\tau )$ of seismic
noise. The main conclusion is that, in the semi-classical  regime, even
if the source noise is not homogeneous, the field correlation
is very close to the Green's function; in many cases, there is only a
prefactor
which I computed and which introduces no  phase shift.
This prefactor vanishes if the support of the source noise does not
meet the rays from $B$ to $A$.

 Many other ideas and applications  remains to be exploited:

Is it possible to use the previous tools in order to get
informations on the source noise?
Can I extend the previous calculus to the case where  the
source noise is located on a surface?
Can I do something similarly in other regimes
of propagation, in particular  in non-smooth media? 
Can I get applications of the general formula to monitoring?

\bibliographystyle{plain}

\end{document}